\documentclass[conference]{IEEEtran}

\usepackage{cite}
\usepackage{amsmath}         
\usepackage{amssymb}         
\usepackage{amsfonts}
\usepackage{textcomp}
\usepackage{graphicx}        
\usepackage{url}             
\usepackage{algorithm}       
\usepackage{algorithmic}     

\usepackage{multirow}        
\usepackage{booktabs}        
\usepackage{tabularx}        
\usepackage{array}           

\usepackage{subfigure}       
\usepackage{caption}         
\usepackage{float}           

\usepackage{hyperref}        
\usepackage{cleveref}        
\usepackage{color}           
\usepackage{xcolor}          
\usepackage{listings}

\definecolor{diffremoved}{rgb}{1.0,0.6,0.6}
\definecolor{diffadded}{rgb}{0.6,1.0,0.6}

\IEEEoverridecommandlockouts

\newcolumntype{C}[1]{>{\centering\arraybackslash}p{#1}}

\begin{document}

\title{An AST-guided LLM Approach for SVRF Code Synthesis}


\author{
\begin{tabular}{c@{\hspace{8em}}c}
    1\textsuperscript{st} Abanoub E. Abdelmalak & 2\textsuperscript{nd} Mohamed A. Elsayed \\
    \textit{Calibre Yield Enhancement Services} & \textit{Calibre Yield Enhancement Services} \\
    \textit{Siemens EDA} & \textit{Siemens EDA} \\
    Cairo, Egypt & Munich, Germany \\
    abanoub.elkess@siemens.com & mohamed.a.elsayed@siemens.com \\[1em]
    3\textsuperscript{rd} Ilhami Torunoglu & 4\textsuperscript{th} David Abercrombie \\
    \textit{Calibre Research \& Development} & \textit{Calibre Product Management} \\
    \textit{Siemens EDA} & \textit{Siemens EDA} \\
    CA, USA & NC, USA \\
    ilhami.torunoglu@siemens.com & david.abercrombie@siemens.com
\end{tabular}
}

\maketitle

\begin{abstract}
Standard Verification Rule Format (SVRF) is essential for semiconductor applications like Design Rule Check (DRC), Layout Versus Schematic (LVS), and Optical Proximity Correction (OPC) and it faces challenges as advancing nodes create complex design rules that renders traditional SVRF development ineffective and highlight an expertise gap. This paper introduces a novel methodology integrating Abstract Syntax Tree (AST) embedding and Retrieval-Augmented Generation (RAG) for enhanced SVRF code synthesis, ensuring semantic accuracy and error minimization through structural validation with domain-specific insights for precise code generation.

We evaluate different T5-based models and propose an innovative SVRF-specific scoring framework that complements standard metrics like BLEU and ROUGE-L. In our approach, AST provides rigorous structural validation, while RAG infuses relevant domain knowledge, effectively enhancing the code generation workflow.


Testing on a comprehensive benchmark of 740 DRC rule implementations, our methodology demonstrates up to a 40\% improvement in code generation accuracy compared to basic text-based fine-tuning process. This fusion of industry expertise with advanced coding strategies not only optimizes SVRF development under limited dataset constraints but also creates a more intuitive and efficient coding environment. Consequently, users can rapidly iterate through design cycles, reduce manual error correction, and significantly improve overall productivity.

\end{abstract}


\section{\textbf{Introduction}}
The advancement in semiconductor technology toward smaller nodes has introduced unprecedented complexity into process development, necessitating precise methodologies. This is particularly evident in domains such as Optical Proximity Correction (OPC), where accurate recipe creation ensures fabricated designs adhere to stringent process margins. This complexity manifests in thousands of intricate design rules and device definitions that designers must follow. These are translated into rule-decks for critical verification steps managed by Electronic Design Automation (EDA) tools, such as Design Rule Checking (DRC) and Layout Versus Schematic (LVS) checking. The proprietary Standard Verification Rule Format (SVRF) is crucial in developing these rule decks for design compliance verification and OPC recipes for manufacturing. However, the industry's rapid growth has created an urgent demand for engineers with deep process understanding and robust development capabilities. The extensive training required, often spanning years, leads to a significant knowledge gap that impacts development efficiency and scalability.

Large Language Models (LLMs) demonstrate potential for automating code generation~\cite{Brown2020LanguageMA} for open-source languages. While LLMs show promise in specialized domains~\cite{Zheng2023ASO}, applying standard pre-trained LLMs to generate SVRF code often results in high hallucination rates with syntactically and semantically invalid outputs. This limitation stems from the semiconductor industry's constrained nature and the scarcity of public information about development methodologies and objectives, highlighting the need for specialized LLM approaches integrated with domain-specific safeguards.

A significant challenge lies in the absence of general SVRF datasets, rooted in the proprietary nature of the language and the semiconductor manufacturing companies that write the code. Since SVRF rule-decks represent semiconductor manufacturing processes, access remains restricted to a limited set of specialized engineers.

We propose an Abstract Syntax Tree (AST) guided methodology for fine-tuning pre-trained LLMs, capturing SVRF's nuances without extensive training data. This approach is enhanced by a Retrieval-Augmented Generation (RAG) component that grounds outputs in verified patterns and documentation, mappable to specific semiconductor manufacturing processes or methodologies, thereby improving solutions across SVRF sub-domains.

To validate our methodology, we apply it to DRC, developing rule decks for foundry design rule compliance. Our dataset, generated using internal knowledge without links to real-world foundry data, enables us to evaluate LLM performance using natural language prompts for code generation. We compare AST-guided models against purely text-based fine-tuning, aiming to generalize the methodology across various SVRF domains beyond DRC.

Our results demonstrate significant improvements in SVRF code accuracy while reducing hallucinations and maintaining logical consistency across complex layer interactions. This approach supersedes traditional template-based code generation methodologies, enabling flexible, intelligent models with diverse use-cases. We envision its role in agentic workflows incorporating SVRF knowledge with multi-agent capabilities, such as vision for layout inspection and tool integration, benefiting both developers and end-users.


\section{\textbf{Methodology}}
\label{sec:methodology}

\begin{figure}
    \centering
    \includegraphics[width=0.35\textwidth]{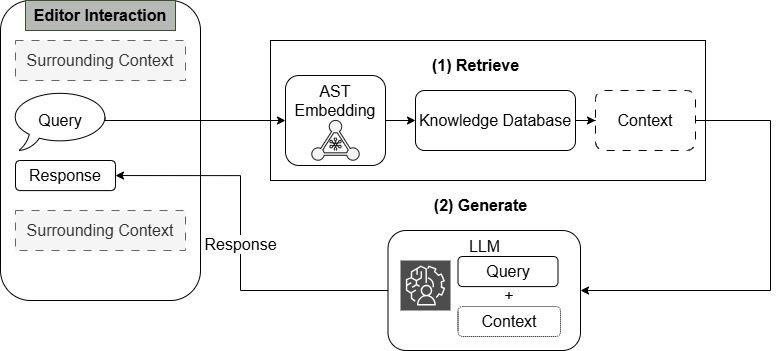}
    \caption{System Design Overview}
    \label{fig:system-overview}
\end{figure}

Our approach to SVRF code synthesis combines the reasoning capabilities of LLMs with domain-specific validation mechanisms. The system architecture, illustrated in Figure \ref{fig:system-overview}, consists of two primary components that work in concert to ensure accurate and efficient context-aware code generation. (1) the retrieval block, where we match the given user input query with the best possible matching given the current user context and the knowledge database. (2) the generation block, where we use our model to generate the code that fits the current query.

\subsection{\textbf{Abstract Syntax Tree (AST)}} 
\label{subsec:ast}

\subsubsection{AST Construction and Preprocessing:}
Abstract Syntax Trees (ASTs) are foundational, representing SVRF code's hierarchical syntactic structure. Each node corresponds to a source code construct, and ASTs serve critical functions in our preprocessing pipeline and LLM guidance, including: syntactic validation, semantic analysis (capturing layer definitions, control flow, command relationships for better LLM context), and LLM integration (providing structured training data and a basis for structural correctness evaluation).

We define SVRF's core components (e.g., "COMMAND", "LAYERS") using ANTLR~\cite{parr2013definitive} grammar for precise AST construction~\cite{aho2007compilers}, ensuring generated code can be assessed for integrity~\cite{baxter2004ast} (see Appendix~\ref{sec:appendix_ast_details} for an AST mapping example). The AST structure enables comprehensive code analysis: command recognition identifies operations and components; layer parsing extracts names and maps relationships; condition handling manages constraints; and option organization provides a framework for parameters. This ensures components are validated and coherently structured.

For optimal LLM consumption, initial ANTLR-parsed SVRF examples are further preprocessed by:
\begin{itemize}
    \item Streamlining the parse tree into a more abstract AST (removing redundancies, standardizing node types).
    \item Serializing this AST into a linearized, bracketed string (e.g., \texttt{(COMMAND (OPTION val) ...)}) via depth-first traversal, preserving hierarchy for LLM tokenization.
\end{itemize}

\subsubsection{AST-Guided LLM Integration and Fine-tuning}
\label{sec:ast_fine_tuning_integration_w_llm}

Our AST-guided approach enhances CodeT5's capabilities~\cite{wang-etal-2021-codet5} by incorporating structural and semantic insights from AST representation throughout both training and inference phases. This integration is crucial for accurate SVRF code generation with limited data, as it efficiently encodes SVRF syntax and logic independent of specific values, mitigating extensive data augmentation needs and reducing dataset size and computational demands.

During fine-tuning, T5 models are trained to translate Natural Language (NL) descriptions into SVRF code strings using a specialized AST-weighted loss function. Candidate and ground-truth SVRF are parsed into ASTs, and the loss function compares these structures, penalizing discrepancies based on their significance (e.g., higher penalties for errors in commands or layers versus minor options). This structural feedback, guided by weights reflecting SVRF grammar priorities (detailed in Appendix~\ref{appendix:metrics}), helps the model learn syntactic and semantic rules more effectively than standard losses, bridging semantic understanding with syntactic requirements.

During inference, learned structural knowledge implicitly guides decoding towards valid SVRF. Correctness is further enhanced by lightweight ANTLR grammar parsing during beam search or post-generation to penalize/discard malformed snippets. Generated SVRF is parsed into an AST and validated syntactically, with errors flagged or potentially corrected.

\begin{figure}[hbp!]
    \centering
    \includegraphics[width=1\linewidth]{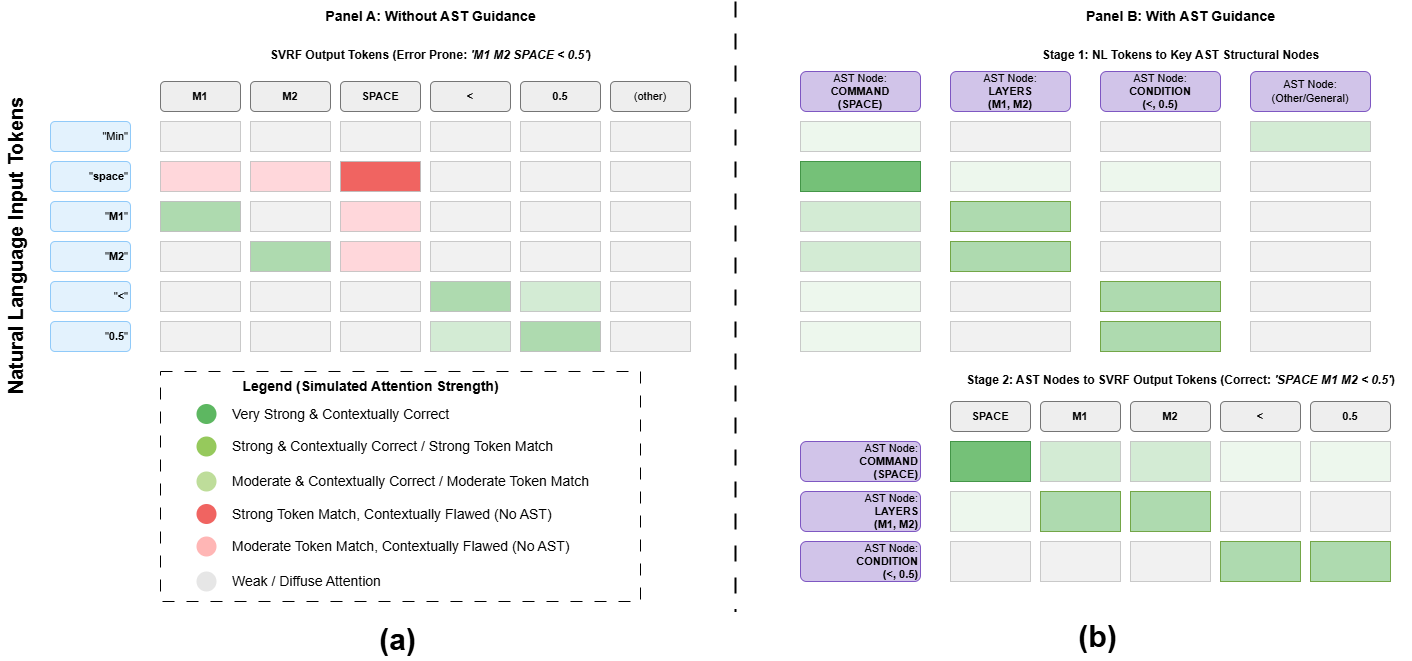}
    \caption{Simulated Token-Level Attention Comparison: (a) Without AST (b) With AST Guidance.}
    \label{fig:heatmap_context_comparison}
\end{figure}

Figure~\ref{fig:heatmap_context_comparison} illustrates this contextual understanding. Without AST guidance (Panel a), LLMs may correlate literal NL tokens to SVRF counterparts but miss structural rules, leading to errors like misplacing "SPACE" after layers. With AST guidance (Panel b), NL tokens correctly attend to corresponding AST structural nodes (e.g., NL "space" to \texttt{COMMAND (SPACE)} AST node), enabling proper structural ordering and syntactically correct SVRF generation.

Key advantages include enhanced syntactic and semantic alignment through hierarchical code representation, reduced errors via structural validation, efficient learning from limited data, and better pattern recognition of command hierarchies for coherent generation.

\subsection{\textbf{Model Architecture Selection}}
\label{sec:model_architecture}
We selected the T5 architecture family for our experiments due to its encoder-decoder architecture, which has shown superior performance in structured generation tasks~\cite{raffel2020exploring}. Unlike autoregressive models like GPT, T5's encoder-decoder structure provides several key advantages for SVRF code synthesis: bidirectional context understanding in the encoder for capturing complex design rule relationships, structured decoding process that better maintains syntactic consistency, and enhanced ability to map between different formats (natural language to code) through parallel attention mechanisms~\cite{vaswani2017attention}.

For our implementation, we utilized three variants of T5-based models:
\begin{itemize}
    \item T5-base~\cite{raffel2020exploring}: The foundational model (220M parameters). Specialized for code generation.
    \item Flan-T5-base~\cite{chung2022scaling}: An instruction-tuned variant (250M parameters)
    \item CodeT5-base~\cite{wang-etal-2021-codet5}: Code-specific pre-trained model (220M parameters)
\end{itemize}

Importantly, we maintained the models' original tokenizers without custom modifications, demonstrating the adaptability of standard pre-trained vocabularies to SVRF syntax.

\subsection{\textbf{Retrieval-Augmented Generation Workflow}}

At the core of our methodology is a RAG workflow, which is designed to optimize the SVRF code generation process by leveraging an extensive knowledge base. This system enhances the code generation capability LLMs by incorporating contextual and domain-specific information.

\subsubsection{Contextual Retrieval Mechanism}

Our RAG mechanism is tailored specifically for SVRF code patterns. Unlike conventional RAG approaches, our system integrates semiconductor process knowledge and tool-specific syntax patterns. It maintains a curated database of verified SVRF code snippets that are indexed using not only syntactic characteristics but also their associated physical verification intents. When tasked with implementing a new DRC rule, the system performs an analysis of the input specification to identify key verification requirements.

\subsubsection{Knowledge Graph Integration Through AST}

The retrieval process is further enhanced by a knowledge graph that encapsulates extensive domain-specific expertise. This graph captures the relationships between various semiconductor processes and their corresponding SVRF code implementations, enabling the system to rank candidate SVRF patterns based on both syntactic similarity and semantic relevance. In addition, by mapping the internal dataset to its equivalent AST representation, our approach leverages structural context to focus on the essential code elements—such as commands and options—thereby reducing dependency on user-specific inputs like layer names or values.


\subsubsection{Prompt Enhancement and Code Generation}

Once relevant SVRF patterns are retrieved, they are used to formulate an enhanced prompt for the LLM. This retrieval-informed prompt provides structured clues and rich contextual information, substantially improving the model's understanding and response to user queries. As a result, the generated code snippets are not only syntactically and structurally accurate but also closely aligned with the intended verification objectives.

By integrating retrieval augmentation with advanced generation techniques, our workflow offers a sophisticated solution for SVRF code development. This approach effectively combines powerful data-driven insights with intuitive code synthesis, leading to improved accuracy and efficiency in design rule checking and verification processes.

\section{\textbf{Experimental Evaluation and Analysis}}
\label{sec:experiments}

In this section, we present a systematic evaluation of our approach through a series of progressive experiments that validate our initial hypothesis. First, we introduce the dataset, the selected pre-trained LLMs, and the evaluation metrics. We then detail our experimental pipeline, analyzing the results at each phase as summarized in Table~\ref{tab:comprehensive_results} and analyzed in the results subection.

\subsection{\textbf{Dataset Definition}}
\label{sec:dataset_defintion}

Our experimental dataset is derived from an internal Design Rule Checking (DRC) knowledge base, initially comprising 400 paired examples of natural language descriptions and their corresponding SVRF code implementations. Through data augmentation techniques using our internal LLM tools, we expanded this to 741 diverse examples.

\subsubsection{Data Structure and Representation}

Each example in our dataset consists of two main components:
\begin{itemize}
    \item Input: NL description of design rules.
    {\small
    \begin{verbatim}
"Minimum spacing between METAL1 and METAL2
     layers should not be less than 0.5um"
    \end{verbatim}
    }
    
    \item Output: Corresponding SVRF code implementation.
    {\small
    \begin{verbatim}
SPACE_CMD METAL1 METAL2 >= 0.5 READ ALL {
    REPORT "Spacing violation detected"
}
    \end{verbatim}
    }
\end{itemize}

The dataset construction process involved:
\begin{itemize}
    \item Initial Collection: 400 curated description-code pairs from internal DRC knowledge
    \item Data Augmentation: LLM-based generation of variations maintaining semantic validity
    \item Quality Assurance: Verification of generated examples by domain experts
\end{itemize}

\subsubsection{Exploratory Data Analysis}

The final dataset of 741 examples exhibits the following complexity distribution:

{\footnotesize
\begin{table}[htp]
\centering
\caption{Distribution of Rules Across Complexity Categories and Dataset Splits}
\label{tab:dataset_distribution}
\begin{tabular}{lcccc}
\hline
\textbf{Complexity} & \textbf{Count} & \textbf{Train} & \textbf{Val} & \textbf{Test} \\
& & \textbf{(80\%)} & \textbf{(10\%)} & \textbf{(10\%)} \\
\hline
Simple Rules & 241 & 193 & 24 & 24 \\
Moderate Rules & 347 & 278 & 35 & 34 \\
Complex Rules & 153 & 122 & 15 & 16 \\
\hline
\textbf{Total} & 741 & 593 & 74 & 74 \\
\hline
\end{tabular}
\end{table}
}

The complexity categories are defined based on:
\begin{itemize}
    \item Simple Rules (32.5\%): Basic layer operations, single command structures, and minimal option parameters
    \item Moderate Rules (46.8\%): Multi-layer interactions, combined operations, and standard option configurations
    \item Complex Rules (20.7\%): Nested operations, multiple layer dependencies, and advanced option combinations
\end{itemize}

The dataset's quality is ensured through careful preservation of verification intent and structural validity, while maintaining diverse patterns across different complexity levels. Using a standard 80-10-10 split ratio, we divided the 741 examples into training (593), validation (74), and testing (74) sets. This organization supports comprehensive code analysis and generation, enabling proper handling of commands, parameters, and configurations while maintaining systematic error detection throughout the development pipeline.



\subsection{\textbf{Baseline Models}}

To evaluate the effectiveness of our AST-guided fine-tuning approach, we implement several baseline models, utilizing different pre-trained language models as foundations:

\begin{itemize}
    \item Pre-trained Local Models: As mentioned in Section~\ref{sec:model_architecture}, we evaluate three pre-trained transformer models: CodeT5-base, Flan-T5-base, and T5-base as our baseline architecture.
    \item Large state-of the art LLMs: Claude sonnet 3.5 is used through Direct prompting. Basic SVRF documentation is added as part of the prompt context window to guide the model.
\end{itemize}

These baselines were chosen to address specific research questions:
\begin{itemize}
    \item How much does structural awareness (AST guidance) improve over standard fine-tuning across different pre-trained models?
    \item Which pre-trained model architecture best suits SVRF code generation?
    \item What is the relative contribution of retrieval versus structural guidance?
\end{itemize}

\begin{table*}[htp]
    \caption{Comprehensive Performance Evaluation of Models Across Different Phases}
    \centering
    \setlength{\tabcolsep}{4pt}
    \begin{tabular}{ll|cccc|cccc|cccc|cccc}
    \hline
    \multirow{2}{*}{\textbf{Model}} & \multirow{2}{*}{\textbf{AST}} & 
    \multicolumn{4}{c|}{\textbf{Zero-shot}} & 
    \multicolumn{4}{c|}{\textbf{Training}} & 
    \multicolumn{4}{c|}{\textbf{Validation}} & 
    \multicolumn{4}{c}{\textbf{Testing}} \\
    & & Loss & Acc & BLEU & R-L & Loss & Acc & BLEU & R-L & Loss & Acc & BLEU & R-L & Loss & Acc & BLEU & R-L \\
    \hline
    \multirow{2}{*}{T5} & w/o & 8.603 & 0.000 & 0.085 & 0.296 & 0.005 & 87.495 & 0.994 & 0.997 & 0.913 & 39.083 & 0.598 & 0.708 & 0.761 & 50.289 & 0.702 & 0.780 \\
    & w/ & 8.603 & 0.000 & 0.085 & 0.296 & 0.034 & 85.434 & 0.978 & 0.987 & 0.247 & 51.796 & 0.776 & 0.833 & 0.237 & 56.042 & 0.796 & 0.865 \\
    \hline
    \multirow{2}{*}{Flan-T5} & w/o & 4.398 & 0.000 & 0.018 & 0.157 & 0.006 & 84.937 & 0.981 & 0.987 & 0.751 & 37.271 & 0.635 & 0.739 & 0.792 & 46.407 & 0.695 & 0.777 \\
    & w/ & 4.398 & 0.000 & 0.018 & 0.157 & 0.040 & 86.808 & 0.987 & 0.993 & 0.279 & 51.519 & 0.785 & 0.861 & 0.234 & 58.947 & 0.837 & 0.885 \\
    \hline
    \multirow{2}{*}{CodeT5} & w/o & 8.096 & 0.000 & 0.002 & 0.096 & 0.012 & 86.729 & 0.989 & 0.994 & 0.519 & 50.995 & 0.725 & 0.801 & 0.608 & 57.211 & 0.763 & 0.828 \\
    & w/ & 8.096 & 0.000 & 0.002 & 0.096 & 0.005 & 86.003 & 0.992 & 0.995 & 0.175 & 63.796 & 0.876 & 0.916 & 0.220 & 62.879 & 0.840 & 0.898 \\
    \hline
    \end{tabular}
    \label{tab:comprehensive_results}
    \raggedright
    \scriptsize Note: "w/" and "w/o": with/without AST guidance. R-L: ROUGE-L score. Acc: AST Weighted Accuracy (\%). Training epochs: 20. Training time: \~6hrs w/o AST and \~8hrs w AST.
\end{table*}

\subsection{\textbf{Evaluation Metrics}}
\label{sec:evaluation_metrics}
To comprehensively assess model performance across different aspects of SVRF code generation, we employ multiple complementary metrics. These include traditional metrics (Loss, BLEU, and ROUGE-L scores) and a novel AST-weighted accuracy scoring mechanism specifically designed for SVRF's unique characteristics. The complete mathematical formulations and detailed descriptions of these metrics are provided in Appendix~\ref{appendix:metrics}. 

For our analysis, we focus on four key metrics (detailed in Appendix~\ref{appendix:metrics}): Loss Score, measuring token-level prediction accuracy; BLEU Score (Eq.~\eqref{eq:bleu}), evaluating similarity to reference implementations; ROUGE-L Score (Eq.~\eqref{eq:rouge-l}), assessing structural similarity and fluency; and our proposed AST-Weighted Accuracy (Eq.~\eqref{eq:ast_accuracy}), which accounts for SVRF-specific characteristics.


While BLEU and ROUGE-L offer valuable insights into lexical similarity and fluency, they are primarily n-gram based and may not fully capture the structural and semantic correctness crucial for a domain-specific language like SVRF. SVRF's non-sequential command ordering, the critical significance of layer sequence, and the hierarchical nature of its commands and options necessitate a more nuanced evaluation. Our AST-Weighted Accuracy is specifically designed to address these characteristics by dissecting the generated code into its core structural components and evaluating their correctness with differential importance, as detailed in Appendix \ref{appendix:metrics}.

This combination of metrics ensures our evaluation captures both general code generation quality and SVRF-specific requirements. The AST-weighted scoring, detailed in Appendix~\ref{appendix:metrics}, is particularly important as it accounts for SVRF's non-sequential nature and the critical importance of layer ordering in the generated code.





\subsection{\textbf{Experimental Pipeline}}

Our evaluation follows a three-phase approach to systematically assess model performance and the impact of AST guidance.

\subsubsection{\textbf{Phase 1: Baseline Performance}}
Initial zero-shot evaluation reveals significant challenges in SVRF code generation across all models. Despite their sophisticated pre-training, models achieve 0\% accuracy with high loss values (T5: 8.603, Flan-T5: 4.398, CodeT5: 8.096). Flan-T5 demonstrates the lowest initial loss, while T5 shows marginally better BLEU (0.085) and ROUGE-L (0.296) scores, indicating limited transfer of pre-trained capabilities to SVRF generation.

\subsubsection{\textbf{Phase 2: Standard Fine-tuning}}
Traditional supervised fine-tuning yields substantial improvements, with all models achieving training accuracies above 84\%. CodeT5 demonstrates particularly strong training performance (86.729\% accuracy, 0.989 BLEU, 0.994 ROUGE-L), followed by T5 and Flan-T5. In validation, CodeT5 maintains its lead with 50.995\% accuracy, suggesting better generalization capabilities even without structural guidance.

\subsubsection{\textbf{Phase 3: AST-Guided Fine-tuning}}
The integration of AST-guided fine-tuning demonstrates statistically significant performance improvements across all transformer architectures, with CodeT5 exhibiting optimal convergence characteristics. The model achieves a minimal cross-entropy loss of 0.005 during training while maintaining 86.003\% AST-weighted accuracy, indicating efficient parameter optimization. In validation, the model demonstrates superior performance metrics with a 63.796\% AST-weighted accuracy, coupled with high sequence-based similarity scores (BLEU: 0.876, ROUGE-L: 0.916), suggesting robust structural learning. The minimal performance degradation in testing (62.879\% accuracy) indicates effective mitigation of overfitting, with only a 0.917 percentage point drop from validation to testing, demonstrating robust generalization across the latent space of SVRF code structures.

\subsection{\textbf{Results Summary}}

\subsubsection{\textbf{Analysis}}
Our experimental results demonstrate the substantial impact of AST-guided fine-tuning on SVRF code generation. CodeT5 emerges as the clear leader, showing exceptional performance across all phases. With AST guidance, it achieves remarkable validation metrics (63.796\% accuracy, 0.876 BLEU, 0.916 ROUGE-L) and maintains strong performance in testing (62.879\% accuracy, 0.840 BLEU, 0.898 ROUGE-L).

\textbf{Overfitting Mitigation through AST:}
A critical observation from our experiments is the role of AST guidance in addressing overfitting issues. Models trained without AST show clear signs of overfitting, with significant performance gaps between training and testing phases (e.g., CodeT5 without AST: 86.729\% training accuracy vs 57.211\% testing accuracy, a 29.518\% drop). The AST-guided approach substantially reduces this gap (86.003\% to 62.879\%, a 23.124\% drop) by embedding structural knowledge that helps the model generalize better.

This improved generalization can be attributed to several factors:
\begin{itemize}
    \item \textbf{Structural Regularization}: AST guidance acts as an implicit regularizer by enforcing syntactic constraints during training
    \item \textbf{Knowledge Embedding}: Instead of merely memorizing patterns, models learn meaningful code structures through AST representations
    \item \textbf{Consistent Performance}: Smaller validation-testing gaps (63.796\% to 62.879\%) indicate more robust learning
\end{itemize}

\begin{figure}[htp]
    \centering
    \includegraphics[width=0.9\linewidth]{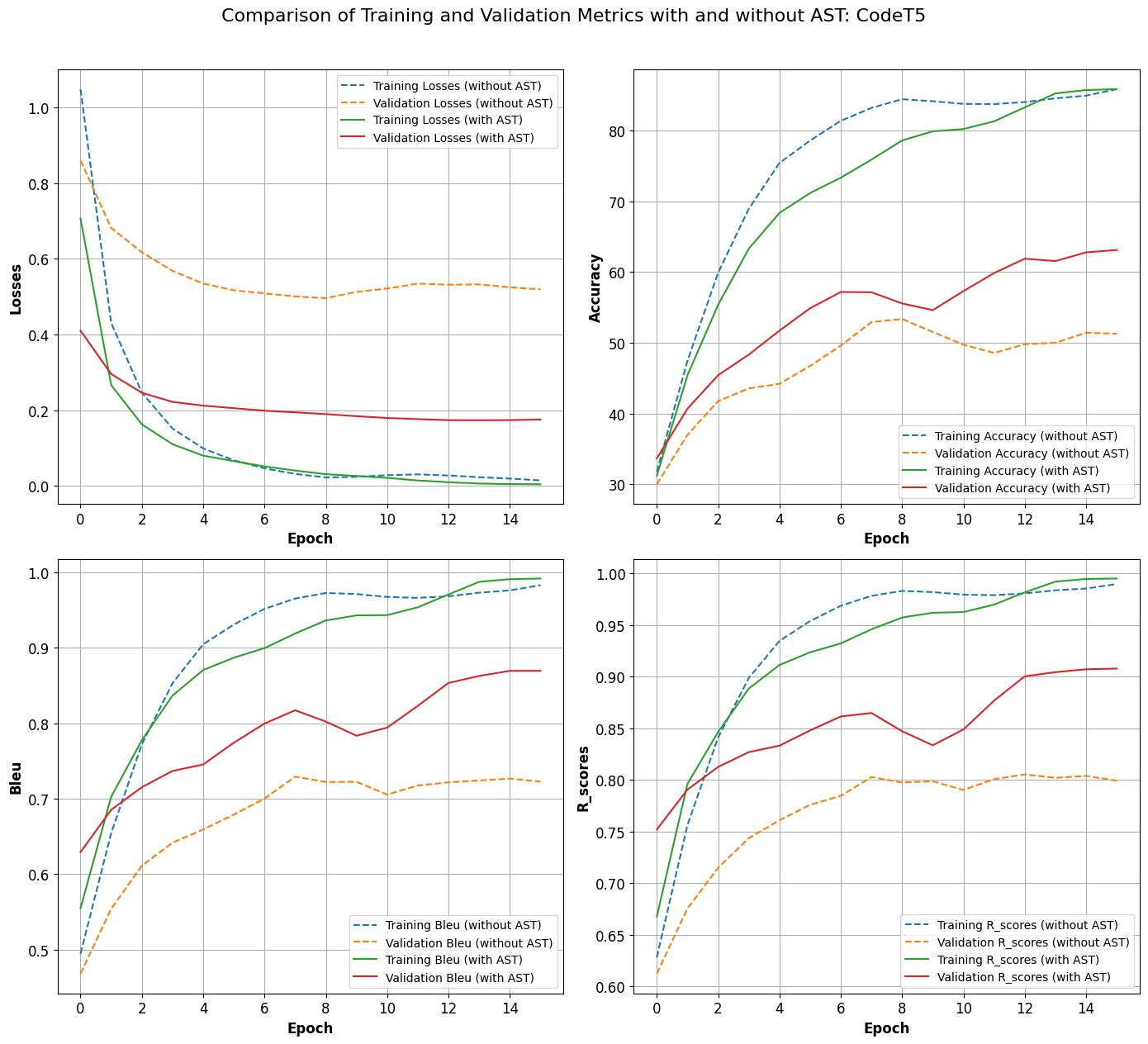}
    \caption{Code-T5: Learning Curves with/without AST}
    \label{fig:learning_curves}
\end{figure}

The learning curves (Figure~\ref{fig:learning_curves}) further support this observation, showing more stable validation metrics and reduced oscillation in the AST-guided approach, indicating better generalization capabilities without sacrificing model capacity.

\subsubsection{\textbf{Model Performance Comparison}}
Each model demonstrates distinct characteristics in handling SVRF code generation. CodeT5, leveraging its code-specific pre-training, shows superior performance with AST guidance, achieving the highest scores across all metrics and phases. The model's validation-to-testing performance remains notably consistent, suggesting robust generalization capabilities. Flan-T5 exhibits significant improvement with AST integration, showing a \~40\% relative increase in validation accuracy and ~27\% in testing accuracy. While T5 shows modest improvements with AST guidance (accuracy increase from 50.289\% to 56.042\%), its performance consistently trails behind both CodeT5 and Flan-T5. Notably, all models demonstrate reduced validation-testing performance gaps with AST guidance, indicating improved generalization capabilities across different architectures.

The detailed examination of learning dynamics, including loss convergence characteristics and metric evolution patterns, is provided in Appendix~\ref{appendix:learning_dynamics} due to space constraints.

\section{\textbf{Application Layer Integration}}
\label{sec:appLayer}
Our methodology is further extended through practical integration within the development environment, enabling a copilot-like experience. The system leverages both the RAG framework and AST-guided code synthesis to deliver context-aware, real-time code suggestions.

\subsection{RAG Integration}
Our system enhances code generation through a RAG framework~\cite{lewis2020retrieval} that leverages contextual information from multiple sources. This additional context includes workspace-specific patterns, historical implementations, and environment configurations. The RAG engine continuously indexes and analyzes these patterns~\cite{krishna2022rag}, building an adaptive knowledge base that evolves with usage. By combining this rich contextual information with our fine-tuned models, the system generates more accurate and environment-aware code suggestions.

\subsection{Editor Environment Integration}
We leverage the development environment to seamlessly connect our RAG infrastructure with the user's workspace. The system begins by analyzing the current code context, capturing the immediate development environment and active coding patterns. This real-time analysis provides crucial insights into the developer's current task and coding style.

The RAG knowledge base then retrieves relevant patterns and examples from its indexed repository, considering both historical implementations and current best practices. This retrieved context is carefully weighted and filtered to ensure relevance to the current development scenario.

The system combines both the immediate coding context and the retrieved knowledge to generate predictions. This dual-context approach enables more nuanced and accurate suggestions, taking into account both the specific requirements of the current task and broader coding patterns. Finally, the system delivers these enhanced suggestions in real-time, maintaining a natural flow within the development process while significantly improving the quality and relevance of generated code~\cite{weigelt2020workflow}.

\section{\textbf{Conclusion and Future Work}}
\label{sec:conclusion}
In this paper, We identified a significant gap in the current LLM-based code generation approaches for specialized domains such as SVRF code synthesis. To address this gap, We proposed a methodology through a combination of AST generation and supervised fine-tuning of pre-trained models. Based on our experiments, around 40\% enhancement in code generation is observed when using an AST approach for fine-tuning versus a standard text-based fine-tuning.

The experimental results demonstrate that AST guidance significantly enhances both model performance and learning efficiency. Our approach enables better generalization, evidenced by the minimal gap between training and validation metrics (0.917 percentage points) and reduced overfitting tendencies. The stable learning progression, characterized by smooth convergence curves and consistent cross-phase performance, indicates that structural information helps the model develop robust understanding of SVRF patterns rather than merely memorizing surface-level features.

However, while AST guidance significantly improves model performance, the validation and testing accuracies (peaking at 63.796\% and 62.879\% respectively with CodeT5) indicate substantial room for improvement. This performance ceiling can be attributed to several key factors:

\begin{itemize}
    \item The inherent complexity of code generation extends beyond structural correctness, requiring deep understanding of operation relationships, precedence, and scope
    \item Current dataset limitations, despite augmentation techniques, may not fully capture the diverse range of possible code structures and their variations
    \item The AST-guided approach, while effective at enforcing structural constraints, could benefit from additional semantic validation mechanisms, such as type checking, scope analysis, and operation compatibility verification
\end{itemize}

Furthermore, we presented an application layer integration that combines a RAG framework with real-time editor environment analysis, offering a copilot-like experience that enhances both the quality and relevance of generated code. This dual-context approach not only streamlines the development process but also reduces manual intervention and error correction.

Looking forward, several key areas deserve further exploration:
\begin{itemize}
    \item Curated data-set collection: Further effort is needed to collect a larger dataset that is more representative of SVRF coding, within the DRC domain as well as other domains
    \item AST-Weighted Loss Function: While our current AST-weighted metric effectively evaluates structural correctness, integrating it directly into the training objective could enhance the model's ability to learn code structures. By designing a differentiable AST-based loss component that penalizes structural mismatches during training, we could guide the model to develop better internal representations of code hierarchies. This could potentially address the current disparity between traditional cross-entropy loss optimization and AST-weighted evaluation metrics, leading to more structurally-coherent code generation.
    \item Enhanced model-tuning methodology: Exploring alternative fine-tuning strategies—such as leveraging Graph Neural Networks to treat ASTs as graphs—may further improve model performance and generalization.
    \item Deeper Application integration: Utilizing the developed model within a larger coding infrastructure with clear and defined features paves the way for a copilot experience, and provides clear value to end users
\end{itemize}

However, it's important to note that the code, tools, and the dataset utilized in this research were developed using an internal proprietary language and are intrinsically tied to sensitive internal information. Consequently, to maintain confidentiality and protect proprietary assets, these materials are not planned for public open-source release.

This paper provides comprehensive implementation details in Section \ref{sec:methodology}, including the AST guided approach (Section \ref{subsec:ast}), model architectures (Section \ref{sec:model_architecture}), evaluation metrics (Section \ref{sec:evaluation_metrics}, Appendix \ref{appendix:metrics}), and experimental setup (Section \ref{sec:experiments}) to enable reproducibility. Implementation questions can be directed to the corresponding author.

In conclusion, we offer a methodology to enhance how SVRF code is developed using LLM infrastructure, which paves the way for fast turn-around-time and higher code quality.

\newpage

\appendix

\subsection{\textbf{AST Mapping Details}}
\label{sec:appendix_ast_details}
For illustration purpose, we use an analogous simplified syntax that maintains the structural essence of SVRF while preserving confidentiality. Note that the following examples use representative commands and structures that parallel actual SVRF syntax without revealing proprietary details.

Consider this basic spacing rule:

{\footnotesize
\begin{verbatim}
    SPACE_CMD METAL1 METAL2 >= 0.5 READ ALL {
        REPORT "Spacing violation detected"
    }
\end{verbatim}
}

This rule is decomposed into the following AST structure:

{\footnotesize
\begin{verbatim}
    <COMMAND>
        <NAME> SPACE_CMD </NAME>
        <LAYERS>
            <LAYER1> METAL1 </LAYER1>
            <LAYER2> METAL2 </LAYER2>
        </LAYERS>
        <CONDITION> >= 0.5 </CONDITION>
        <OPTIONS>
            <MODE> READ ALL </MODE>
            <REPORT> 
                "Spacing violation detected" 
            </REPORT>
        </OPTIONS>
    </COMMAND>
\end{verbatim}
}
This hierarchical representation, while using simplified analogous commands, demonstrates the model's comprehensive capabilities in code analysis and generation. The model exhibits structural awareness by understanding command components and their relationships, implements parameter validation to ensure valid numerical values and operators, maintains proper configuration through options validation, and structures appropriate violation reporting through systematic error handling.

Note: The commands (\texttt{SPACE\_CMD}, \texttt{WIDTH\_CMD}, etc.) and their syntax are simplified representations that parallel the structure of actual SVRF commands while maintaining confidentiality of proprietary syntax.

\subsection{\textbf{Extended Metrics Details}}
\label{appendix:metrics}

\subsubsection{Traditional Metrics}
\begin{itemize}
    \item \textbf{Loss Score}: Cross-entropy loss measuring the model's prediction accuracy at the token level. Lower values indicate better performance, with our models typically ranging from 8.196 (poor) to lower values after fine-tuning.
    \item \textbf{BLEU Score}: Bilingual Evaluation Understudy score evaluating the generated code's similarity to the reference implementation. This metric is particularly useful for assessing partial correctness when exact matches aren't achieved.
        {\small
        \begin{equation} \label{eq:bleu}
            \begin{split}
                BLEU = BP \cdot \exp \Biggl( & \sum_{n=1}^{N} w_{n} \log p_{n} \Biggr)
            \end{split}
        \end{equation}
        where $BP$ is the brevity penalty, $w_n$ are weights, and $p_n$ is the n-gram precision.
        }
        
    \item \textbf{ROUGE-L Score}: Longest Common Subsequence (LCS)-based metric measuring the fluency and structural similarity between generated and reference code.
        {\small
        \begin{equation} \label{eq:rouge-l}
            \begin{split}
               ROUGE-L = \frac{2 \times LCS(X,Y)}{length(X) + length(Y)}
            \end{split}
        \end{equation}
            where $X$ and $Y$ are the reference and generated sequences respectively.
        }
\end{itemize}

\subsubsection{AST-Weighted Scoring}
Given SVRF's unique properties as a non-sequential language where command ordering is flexible but layer ordering is critical, we introduce a novel weighted scoring mechanism:

{\small
\begin{multline}
\label{eq:ast_accuracy}
AST\text{-}Weighted~Accuracy = \frac{1}{N} \sum_{i=1}^{N} ( w_{1} \cdot \mathit{c\_acc}_{i} \\ 
+ w_{2} \cdot \mathit{o\_acc}_{i} + w_{3} \cdot \mathit{l\_acc}_{i} )
\end{multline}
Where:
\begin{itemize}
    \item $\text{c\_acc}_i$: Accuracy of command name and structure
    \item $\text{o\_acc}_i$: Correctness of command options and parameters
    \item $\text{l\_acc}_i$: Accuracy of layer ordering and relationships
    \item $w_1, w_2, w_3$: Weighting factors for each component. These were determined through various experiments along with awareness of the SVRF grammar components priority. These experiments focused on the conceptual "correctness" of the SVRF code from different angels for example: using the correct options for a specific command and the layer ordering.
    \item $N$: Number of examples in the evaluation set
\end{itemize}
}

This weighted approach accounts for SVRF's specific characteristics:
\begin{itemize}
    \item Non-sequential Nature: Command ordering flexibility
    \item Layer Significance: Critical importance of layer ordering
    \item Option Flexibility: Variable ordering of command options
    \item Structural Validity: Emphasis on correct command structure
\end{itemize}

The combination of traditional metrics and our AST-weighted scoring provides a comprehensive evaluation framework that captures both general code generation quality and SVRF-specific structural requirements. This dual approach ensures that our assessment considers both syntactic accuracy and semantic correctness in the context of SVRF's unique characteristics.

\subsection{\textbf{Extended Results Analysis Details}}

\subsubsection{\textbf{Learning Dynamics Analysis}}
\label{appendix:learning_dynamics}

Figure~\ref{fig:learning_curves} presents the learning dynamics of CodeT5 with and without AST guidance across four key metrics: loss, accuracy, BLEU, and ROUGE-L scores. The curves reveal several important patterns:

\textbf{Loss Convergence:}
The AST-guided model exhibits superior loss reduction characteristics throughout the training process. It achieves faster initial convergence in both training and validation phases, demonstrating the effectiveness of structural guidance in accelerating learning. The learning trajectory remains notably stable with minimal fluctuations, contrasting with the more erratic behavior observed in the baseline model. This stability is further emphasized by the significantly lower final validation loss (0.175 compared to 0.519), strongly indicating better generalization capabilities. The consistent and proportional gap maintained between training and validation losses suggests the model achieves an appropriate balance in its capacity, neither underfitting nor overfitting the training data, while effectively leveraging the structural information provided by AST guidance.

\begin{figure}[htp]
    \centering
    \includegraphics[width=0.8\linewidth]{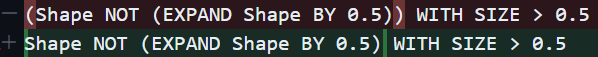}
    \caption{Text-to-Text Comparison: Failure Example}
    \label{fig:failure_example}
\end{figure}

\textbf{AST-Weighted Accuracy Progression:}
The generation quality metrics demonstrate consistent improvements with AST guidance. While traditional metrics like BLEU and ROUGE-L approach near-perfect values during training (improving from 0.725 to 0.876 and 0.801 to 0.916 respectively), the AST-weighted accuracy maintains a more conservative measure, reflecting the structural complexity of the code. This discrepancy is particularly evident in cases where text-based metrics might suggest high similarity despite significant semantic differences. Consider this example in figure \ref{fig:failure_example}, despite differing by only a single opening parenthesis and a parameter name, these expressions have fundamentally different semantic meanings. The background red represents the target code and the background green represents the predicted code.

Text-based metrics would show high similarity scores due to the extensive token overlap, but the AST-weighted accuracy correctly penalizes this generation as structurally incorrect. The missing opening parenthesis fundamentally changes the operator precedence: in the correct version, the NOT operation is applied to the entire expression, while in the generated version, the scope of NOT is ambiguous and would lead to syntax error. This single-character difference, which might appear minor in text-based comparisons, results in completely different AST and, consequently, different semantic meanings. The validation curves exhibit remarkable stability under AST guidance, suggesting more reliable and consistent code generation capabilities that better capture such crucial structural nuances. While both approaches achieve comparable final training performance, their learning trajectories differ markedly, with AST-guided training showing more systematic and controlled progression toward optimal performance, indicating better structural understanding of the code generation task. This is particularly evident in CodeT5's validation performance, where AST guidance improves accuracy from 50.995\% to 63.796\%, maintaining this advantage through testing (57.211\% to 62.879\%). These improvements, while numerically smaller than the near-perfect BLEU and ROUGE-L scores, better reflect the model's true capability in generating structurally valid and semantically correct code.

For future work, we will construct more detailed examples entailing the failures of text-based learning over AST-based learning.

\textbf{BLEU and ROUGE-L Evolution:}
The generation quality metrics reveal consistent and substantial improvement patterns with AST guidance. The approach demonstrates accelerated improvement in both BLEU and ROUGE-L scores, indicating more efficient learning of code generation patterns. Validation performance reaches notably higher plateau levels under AST guidance, with BLEU scores improving from 0.725 to 0.876 and ROUGE-L scores increasing from 0.801 to 0.916. The validation curves maintain remarkable stability with AST guidance, suggesting more reliable and consistent code generation capabilities. While both approaches ultimately achieve comparable training performance, their learning trajectories differ significantly, with AST-guided training exhibiting a more systematic and controlled progression toward optimal performance, indicating enhanced structural understanding of the code generation task.

\textbf{Computational Considerations:}
The integration of AST structures introduces additional computational overhead in both training and inference phases. While the original SVRF code might be relatively concise (e.g., a single-line command), its AST representation significantly expands the token count due to the explicit structural markup. For example, a simple spacing rule of approximately 10 tokens expands to over 30 tokens in its AST form, including structural tags and hierarchical relationships. This expansion necessitates larger maximum sequence lengths during training and inference, requiring 1024 tokens for AST-guided generation compared to 512 tokens used in basic text-based generation. Consequently, training time increases significantly with longer sequences, and memory requirements grow to accommodate the expanded representations. In our implementation, training on an NVIDIA H100 NVL GPU with 95.8GB memory, the AST-guided approach required approximately 8 hours of training time compared to 6 hours for basic text-based fine-tuning, representing a 33\% increase in computational time. This extended training time is accompanied by approximately 40\% higher memory utilization. However, this computational overhead is offset by the improved model performance and reduced need for extensive data augmentation, ultimately providing a more efficient path to robust SVRF code generation.

\subsubsection{\textbf{Relative Accuracy Improvements}}

\begin{figure}[htp]
    \centering
    \includegraphics[width=0.9\linewidth]{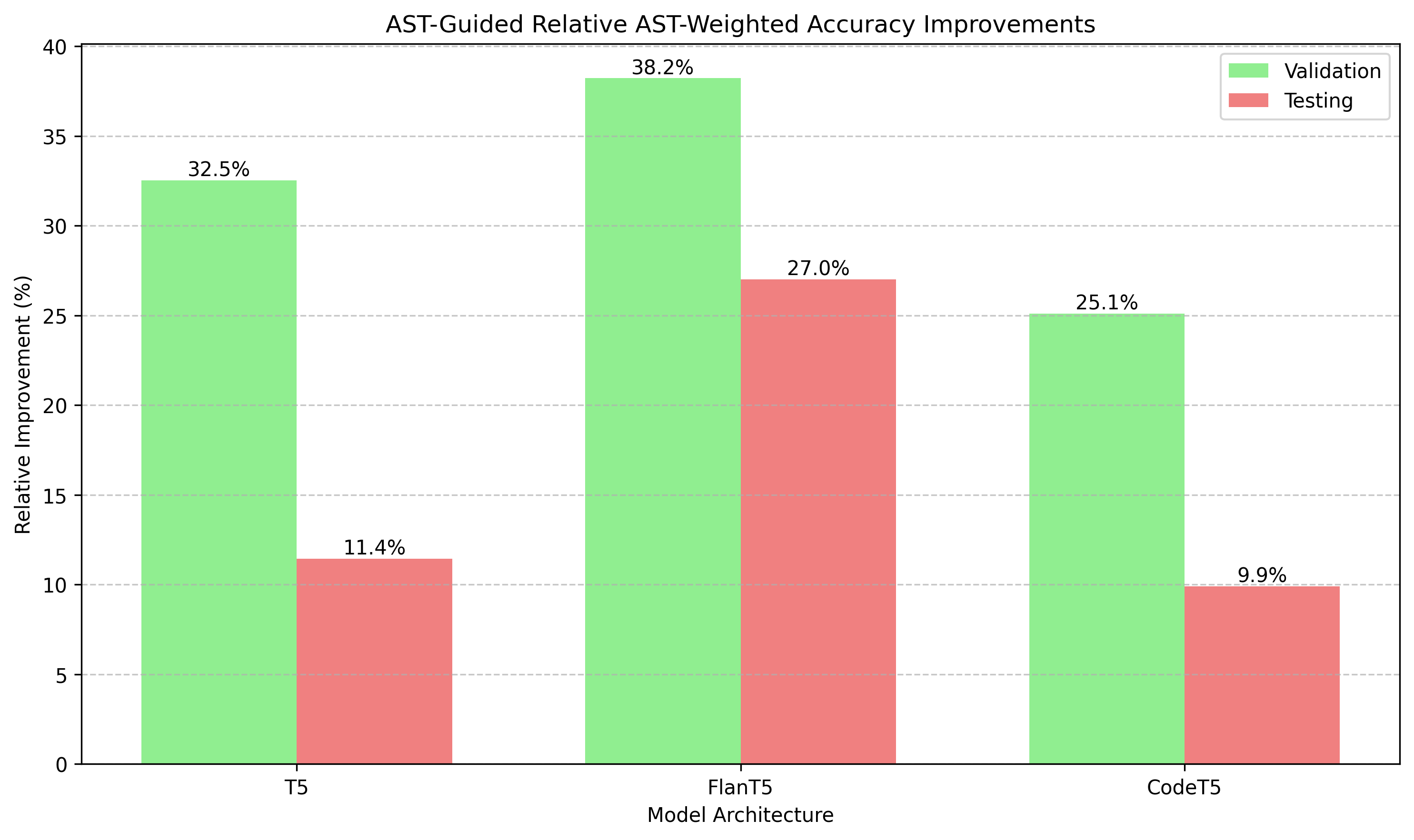}
    \caption{Relative Improvements Comparison with/without AST}
    \label{fig:relative_improvements}
\end{figure}

To better understand the impact of AST guidance, we analyzed the relative improvement in accuracy across different model architectures (Figure \ref{fig:relative_improvements}). This analysis reveals several interesting patterns in how different models respond to AST guidance during validation and testing phases.

FlanT5 demonstrates the most substantial relative improvements, with a 38.2\% increase in validation accuracy and a 27.0\% increase in testing accuracy. This marked improvement suggests that FlanT5's pre-training approach makes it particularly receptive to structural guidance. The consistent improvement across both validation and testing phases (difference of 11.2 percentage points) also indicates robust generalization of the learned structural patterns.

T5 shows the second-highest relative improvement in validation (32.5\%), but this gain diminishes significantly during testing (11.4\%). This substantial drop-off (21.1 percentage points) between validation and testing improvements suggests that while T5 can learn structural patterns effectively, it may be more prone to overfitting when compared to FlanT5.

CodeT5, despite achieving the highest absolute accuracy scores, shows more modest relative improvements: 25.1\% in validation and 9.9\% in testing. This smaller relative gain can be attributed to CodeT5's already strong baseline performance, particularly its inherent understanding of code structures. However, the consistent improvement across phases (difference of 15.2 percentage points) demonstrates that AST guidance still provides meaningful benefits even for code-specialized models.

These patterns suggest that while all models benefit from AST guidance, the magnitude of improvement varies based on the model's architectural strengths and pre-training approach. The more consistent validation-to-testing improvement ratios in FlanT5 and CodeT5 indicate that these architectures may be better suited for maintaining structural learning across different evaluation contexts.

\bibliographystyle{IEEEtran}
\bibliography{bibliography}

\end{document}